\documentclass[journal,twoside]{IEEEtran}
\usepackage{subfigure}
\usepackage{setspace}
\usepackage{amsmath}
\usepackage{amssymb}
\usepackage{amsfonts}
\usepackage{amscd}
\usepackage{mathrsfs}
\usepackage[final]{graphicx}
\usepackage{graphicx}
\usepackage{psfrag}
\usepackage{color}
\usepackage{url}
\newtheorem{theorem}{Theorem}

\newtheorem{definition}{Definition}

\newtheorem{corollary}{Corollary}

\begin{document}

\title{A Non-differential Distributed  Space-Time Coding for Partially-Coherent Cooperative Communication}
\author{Harshan J.~and
        B. Sundar Rajan,~\IEEEmembership{Senior Member,~IEEE}%

\thanks{This work was supported through grants to B.S.~Rajan; partly by the IISc-DRDO program on Advanced Research in Mathematical Engineering, and partly by the Council of Scientific \& Industrial Research (CSIR, India) Research Grant (22(0365)/04/EMR-II). The material in this paper was presented in parts at the IEEE International Conference on Communications (ICC 2008), Beijing, China, May. 19-23, 2008. Harshan J. and B. Sundar Rajan are with the Department of Electrical Communication Engineering, Indian Institute of Science, Bangalore-560012, India. Email:\{harshan,bsrajan\}@ece.iisc.ernet.in.}% <-this % stops a space
\thanks{Manuscript received May 18, 2007; revised January 18, 2008.}}

\markboth{IEEE Transactions on Wireless Communications ,~Vol.~xx, No.~xx, xxxx}{Harshan \MakeLowercase{and} Rajan: A Non-differential Distributed  Space-Time Coding for Partially-Coherent Cooperative Communication}

\maketitle

\begin{abstract}
In a distributed space-time coding scheme, based on the relay channel model, the relay nodes co-operate to linearly process the transmitted signal from the source and forward them to the destination such that the signal at the destination appears as a space time block code. Recently, a code design criteria for achieving full diversity in a partially-coherent environment have been proposed along with codes based on differential encoding and decoding techniques. For such a set up, in this paper, a non-differential encoding technique and construction of distributed space time block codes from unitary matrix groups at the source and a set of diagonal unitary matrices for the relays are proposed. It is shown that, the performance of our scheme is independent of the choice of unitary matrices at the relays. When the group is cyclic, a necessary and sufficient condition on the generator of the cyclic group to achieve full diversity and to minimize the pairwise error probability is proved. Various choices on the generator of cyclic group to reduce the ML decoding complexity at the destination is presented. It is also shown that, at the source, if non-cyclic abelian unitary matrix groups are used, then full-diversity can not be obtained. The presented scheme is also robust to failure of any subset of relay nodes.
\end{abstract}
%%%%%%%%%%%%%%%%%%%%%%%%%%

%%%%%%%%%%%%%%%%%Introduction%%%%%%%%%%%%%%%%%%%%%%%%%%%
\section{Introduction and preliminaries}

Co-operative communication can be based on a relay channel model where a set of distributed antennas belonging to multiple users is exploited for achieving spatial diversity \cite{SEA1}-\cite{JiH1}. In \cite{JiH1}, the idea of Space-Time Coding (STC) devised for point to point co-located multiple antenna systems is applied to a two-hop wireless relay network and is referred as Distributed Space-Time Coding (DSTC). The technique involves a two phase protocol where, in the first phase, the source broadcasts the information to the relays and in the second phase, relays linearly process the received signals and forward them to the destination such that the signal at the destination appears as a Space-Time Block Code (STBC). 

In the above technique, the destination may or may not have the knowledge of channel fade coefficients (i) from the source to the relays and (ii) from the relays to the destination when decoding for the source signal. DSTC when the destination does not have the knowledge of the channels from the source to the relays but has the knowledge of the channels from the relays to itself is called partially-coherent DSTC, which is the subject matter of this paper. For a partially-coherent set up, in \cite{KiR1}, a code design criteria is proposed for achieving full diversity.  A differential encoding and decoding strategy with a class of full diversity achieving codes using cyclic division algebras is also proposed. Inspired by non-coherent MIMO unitary differential modulation \cite{HoS}, differential distributed space time codes have been proposed in \cite{JiJ2}, \cite{OgH1} and \cite{RaR}.

In \cite{KiR1}, it is shown that the coding problem for a partially coherent setup is to distributively design a finite set of unitary matrices such that the matrix obtained from juxtaposing any two distinct matrices from the set must be full rank. In other words, the problem is to design non-intersecting subspaces such that the principal angles between subspaces is as large as possible. However, with the use of differential techniques \cite{KiR1,JiJ2,OgH1, RaR}, the coding problem gets transformed in to a problem of constructing a finite set of unitary matrices such that the difference matrix of any two matrices is full rank. This is a well known design criteria for relay networks with amplify and forward protocol in a coherent detection environment and code constructions based on the above criteria are available in the literature. In this paper, we propose a method to explicitly construct non-intersecting subspaces for a partially-coherent set up. i.e we construct a set of unitary matrices in a distributed way such that the matrix obtained from juxtaposing any two distinct matrices from the set must be full rank.\\
\indent The contributions of this paper can be summarized as follows: (i) for the partially-coherent set up, we introduce a non-differential coding scheme and obtain an expression for ML decoding metric (Theorem \ref{thm1}) for the general class of unitary Distributed Space-Time Block Codes (DSTBCs) (See Definition \ref{udstbc}). (ii) We construct unitary DSTBCs using cyclic unitary matrix groups called Non-differential Cyclic Distributed Space-Time Codes (NCDSTC) (Definition \ref{cdstbc}) and provide a necessary and sufficient condition on its generator such that the NCDSTC is fully diverse (Theorem \ref{thm2}) and the pairwise error probability (PEP) is minimized. (iii) Also, we provide conditions on the choice of the generator of the cyclic group so as to reduce the decoding complexity at the destination. (iv) We show that the proposed scheme is robust to failure of a subset of the relay nodes.

%%%%%%%%%%%%%%%%%%%%%%%%%%ORGANISATIONS%%%%%%%%%%%%%%%%%%%
The remaining part of the paper is organized as follows: In Section \ref{sec2}, along with the signal model, partially-coherent distributed space time coding technique is briefly reviewed and an ML decoding metric for a class of unitary DSTBCs is presented. The scheme of non-differential unitary DSTC from cyclic unitary matrix groups is introduced in Section \ref{sec3} where we provide a necessary and sufficient condition on the generator of the cyclic group to achieve full diversity and to minimize the PEP. In Section \ref{sec4}, an ML decoding metric for the proposed codes is presented along with two reduced decoding complexity ML decoders. Details on the robustness of our scheme to the failure of a subset of the relay nodes is also presented. Possible directions of future work and concluding remarks constitute Section \ref{sec5}.\\
%%%%%%%%%%%%%%%%%%%%NOTATIONS%%%%%%%%%%%%%%%%%%%%%%
\noindent
\textit{Notations:} For a complex matrix $\textbf{X}$, the matrices $\textbf{X}^*$, $\textbf{X}^T$,  $\textbf{X}^{H}$, $|\textbf{X}|$, $\mbox{Re } \textbf{X}$ and $\mbox{Im }\textbf{X}$ denote, respectively, the conjugate, transpose, conjugate transpose, determinant, real part and imaginary part of $\textbf{X}$. 
For a complex matrix $\textbf{Y}$ of the order same as $\textbf{X}$, $\textbf{X} \odot \textbf{Y}$ denotes the Hadamard product of $\textbf{X}$ and $\textbf{Y}$. The $T\times T$ identity matrix is denoted by $\textbf{I}_T$  and $\textbf{O}_T$ denotes the $T$-length vector of zeros.  We use $|x|$ to denote the absolute value of the complex number $x$ and $E \left[x\right]$ to denote the expectation of the random variable $x.$ We write $\textbf{x} \sim \mathcal{CSCG} \left(\mu, \mathbf{\Gamma} \right) $ when $\textbf{x}$ is a circularly symmetric complex Gaussian random vector with mean $\mu$ and covariance matrix $\mathbf{\Gamma}$ and use $j$ for $\sqrt{-1}.$ The set of integers and the set of complex numbers are, respectively, denoted by ${\mathbb Z}$ and ${\mathbb C}.$ 
%%%%%%%%%%%%%%%%Partial Coherent Set up%%%%%%%%%%%%%%%  
\section{Partially Coherent Distributed space time coding}
\label{sec2}

%%%%%%%% Fig.1 %%%%%%%%%%%%%%%
\begin{figure}
\centering
\includegraphics[width= 2.7in]{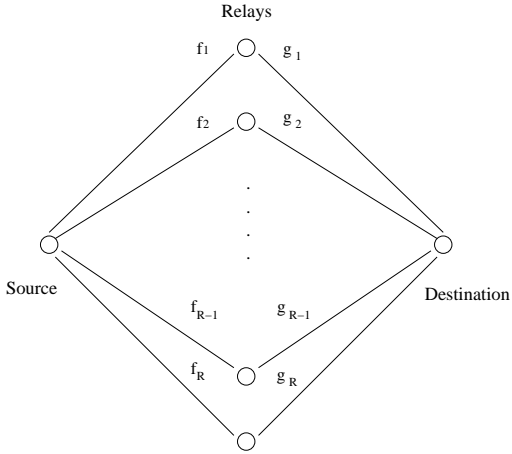}
\caption{Wireless relay network model}
\label{model}
\end{figure}
%%%%%%%%%%%%%%%%%%%%%%%%%%%%

\subsection{Signal model}
The wireless network considered in Figure \ref{model} consists of $R + 2$ nodes each having single antenna which are placed randomly and independently according to some distribution. There is one source node and one destination node. All the other $R$ nodes work as relays. We denote the channel from the source node to the $j$-th relay as $f_{j}$ and the channel from the $j$-th relay to the destination node as $g_{j}$ for $j=1,2, \cdots, R$.
\noindent The following assumptions are made in our system model: (i) all the nodes are subjected to half duplex constraint. (ii) fading coefficients  $f_{j},g_{j}$ are i.i.d $ \mathcal{CSCG} \left(0,1 \right)$ with coherence time interval, $T.$ (iii) all the nodes are synchronized at the symbol level. (iv) Destination knows the fading coefficients $g_{j}$'s but not $f_{j}$'s.\\
%%%%%%%%%%%%%%%%%%%%%First phase of model%%%%%%%%%%%%%
\indent Every transmission from the source to the destination comprises of two phases. In the first phase the source transmits a $T$ length complex vector from the codebook $\mathcal{S}$ = $\left\{ \textbf{s}_{1},\, \textbf{s}_{2},\, \textbf{s}_{3},\, \cdots , \textbf{s}_{L} \right\} $ consisting of information vectors $\textbf{s}_{l} \in \mathbb{C}^{T}$ such that $E\left[\textbf{s}_{l}^{H}\textbf{s}_{l}\right]$ = 1, so that $P_{1}T$ is the average transmit power. In particular, $P_{1}$ is the average transmit power used at the source node for every channel use. When the information vector $\textbf{s}$ is transmitted, the received vector at the $j$-th relay is given by $\textbf{r}_{j} = \sqrt{P_{1}T}f_{j}\textbf{s} + \textbf{n}_{j},  \textit{ j} = 1,2,\cdots, R$. In the second phase, all the relay nodes are scheduled to transmit $T$ length vectors to the destination simultaneously. Each relay is equipped with a fixed  $T\times T$  unitary matrix $\textbf{A}_{j}$ and is allowed to linearly process the received vector. The $j^{th}$ relay is scheduled to transmit $\textbf{t}_{j} = \sqrt{\frac{P_{2}}{(1 + P_{1})}}\textbf{A}_{j}\textbf{r}_{j}$ where $P_{2}$ is the average transmit power used at each relay for every channel use.
\noindent
The vector received at the destination is given by
\begin{equation}\label{bfy}
\textbf{y} = \sum_{j = 1}^{R} (g_{j}\textbf{t}_{j}) + \textbf{w} = \sqrt{\frac{P_{1}P_{2}T}{(1 + P_{1})}}\textbf{S}\textbf{h} + \textbf{n}
\end{equation}

\noindent where $\textbf{w} \sim \mathcal{CSCG} \left(0, \textbf{I}_{T} \right)$ is the additive noise at the destination,
$\textbf{n} = \sqrt{\frac{P_2}{(1 + P_1)}}\left[ \sum_{j=1}^{R} (g_{j}\textbf{A}_{j}\textbf{n}_{j}) \right]  + \textbf{w} \label{bfN}.$ The equivalent channel  \textbf{h} is given by $\textbf{h} = [f_{1}g_{1} ~ f_{2}g_{2} ~ \cdots ~ f_{R}g_{R} ]^T \in \mathbb{C}^{R}$. The codeword matrix $\textbf{S}$ is given by $\textbf{S} = \left[  \textbf{A}_{1}\textbf{s}~~  \textbf{A}_{2}\textbf{s}~~  \cdots ~~ \textbf{A}_{R}\textbf{s} \right] \in {\mathbb C}^{T \times R}.$
%%%%%%%%%%%CODE%%%%%%%%%%%%%%%%%%%%%%%%%%
\begin{definition}
\label{dstbc}
The collection $\mathcal{C} = \left\{ \left[  \textbf{A}_{1}\textbf{s}~~  \textbf{A}_{2}\textbf{s}~~  \cdots ~~ \textbf{A}_{R}\textbf{s} \right]\right\}$ of codeword matrices when $\textbf{s}$ runs over $\mathcal{S}$, is called the Distributed Space-Time Block code (DSTBC).
\end{definition}

%%%%%%%%%%%%%%%%%%%ML DECODER%%%%%%%%%%%%%%%%%%%
\subsection{ML Decoder}
Since the relay matrices $\textbf{A}_{j}$ are unitary, the random vectors $\textbf{w}$ and $\textbf{n}_{j}$ are independent and Gaussian and since $g_{j}$ are known at the receiver, $\textbf{n}$ is a Gaussian random vector with 
\begin{center}
 $E\left[ \textbf{n}\right]$  = $\textbf{0}_{T}$ ~ and
\end{center}
\begin{center}
 $E\left[ \textbf{n}\textbf{n}^{H}\right]$   = $\left(1 + \frac{P_{2}}{(1 + P_{1})}\sum_{j = 1}^{R}(|g_{j}|^{2})\right)\textbf{I}_T.$
\end{center}
Assume that $\textbf{S} \in \mathcal{C}$. When the destination has the knowledge of $g_{j}$'s but not of $f_{j}$'s, ${\textbf y}$ is a Gaussian random vector with
%%%%%%%%%%%%%%%%%%%%%%%Mean%%%%%%%%%%%%%%%%%%%%%%%%%%%%
\begin{center}
$E\left[{\textbf y}|\textbf{S} , g_{j} \right] = \textbf{0}_{T}$ ~ and
\end{center}
%%%%%%%%%%%%%%%%%%%%covariance matrix%%%%%%%%%%%%%%%%%%%%%%%%
\begin{center}
$E\left[ \textbf{yy}^{H}| \textbf{S}, g_{j} \right] = \mathbf{\Sigma}_{y} = \rho \textbf{S} \mathbf{\Sigma}_{h}\textbf{S}^{H} +  \gamma \textbf{I}_T$
\end{center}
where,
\begin{itemize}
\item $\rho = \frac{P_{1}P_{2}T}{P_{1} + 1}$
\item $\gamma = (1 + \frac{P_{2}}{(1 + P_{1})}\sum_{j = 1}^{R}(|g_{j}|^{2}))$ ~ and
\item $\mathbf{\Sigma}_{h} = \mbox{diag}\left( |g_{1}|^{2},\cdots |g_{R}|^{2}\right).$
\end{itemize}
Then, the conditional pdf $P(\textbf{y}| \textbf{S} , g_{j})$ is given by
%%%%%%%%%%%%%%%%%%Distribution of recieved vector%%%%%%%%%%%%%%%%%%%
\begin{equation*}
P(\textbf{y}| \textbf{S} , g_{j}) =\frac{1}{|\mathbf{\Sigma}_{y}|} exp\left( -\left(\textbf y^{H}\mathbf{\Sigma}_{y}^{-1}\textbf y\right)\right) .
\end{equation*}
The partially-coherent ML decoder decodes to a codeword $\hat{\textbf{S}}$ where
\begin{equation}
\label{ML_initial}
\hat{\textbf{S}} = arg\, \max_{\textbf{S} \in \mathcal{C}} P(\textbf{y}|\textbf{S} , g_{j}).\\
\end{equation}

\begin{definition}
\label{udstbc}
\cite{KiR1} A DSTBC is called a unitary DSTBC if every $\textbf{S} \in \mathcal{C}$ in Definition \ref{dstbc} satisfies the condition that $\textbf{S}^{H}\textbf{S} = t \textbf{I}_{R}$, where $t$ is a constant real number independent of the codeword $\textbf{S}$.
\end{definition}
%%%%%%%%%%%%%%%%%%%%%%%%%Theorem 1%%%%%%%%%%%%%%%%%%%
\begin{theorem}
\label{thm1}
For a unitary DSTBC, the partially-coherent ML decoding is given by
\begin{equation}
\label{ML}
arg\, \max_{\textbf{S} \in~\mathcal{C}} \left(\textbf{y}^{H}\textbf{S}\textbf{G}\textbf{S}^{H}\textbf{y}\right), 
\end{equation}
\begin{equation*}
\label{G_matrix}
~  \mbox{ where } ~ \textbf{G}  = \mbox{diag}\left( \beta_{1} ,\cdots \beta_{R} \right) ~ \mbox{ and }
\end{equation*}
$$\beta_{j} = |g_{j}|^{2}\left(|g_{j}|^{2} + \gamma \rho^{-1} \right)^{-1}~ \mbox{for} ~j = 1, \cdots , R.$$

\end{theorem}

\begin{proof} 
Using the well known result, $|\textbf{I} + \textbf{A}\textbf{B}| = |\textbf{I} + \textbf{BA}| $ and the appropriate use of the matrix inversion lemma in \eqref{ML_initial}, the result follows. 
\end{proof}

%%%%%%%Design criteria given by kiran%%%%%%%%%%%%%%%%%%%%%%%%%%%%%%

\subsection{Design criteria}
Chernoff bound on the PEP for the decoder in \eqref{ML_initial} is given in \cite{KiR1} following which a  criteria for designing unitary DSTBCs in order to minimize the PEP is also provided. For achieving a diversity order of $R$, the term $|\textbf{S}_{nm}^{H}\textbf{S}_{nm}|$ has to be non zero for all $m$, $n$ such that $n \neq m,$ where $\textbf{S}_{nm} \in \mathbb{C}^{T \times 2R}$ is obtained by juxtaposing two codewords $\textbf{S}_{n}$ and $\textbf{S}_{m}$ as,
\begin{equation}
\label{concate}
\textbf{S}_{nm}= \left[\textbf{S}_{n} ~~ \textbf{S}_{m} \right].
\end{equation}
Further, for large values of $P$ (where $P = P_{1} + RP_{2}$ is the total power used at all nodes for every channel use), PEP can be minimised by maximising $|\textbf{S}_{nm}^{H}\textbf{S}_{nm}| ~ \mbox{ for  all }  n \neq m.$ A necessary condition for this is $T \geq 2R$. Hence, in the rest of the paper, we consider $T = 2R.$\\
\indent In the following section, we construct codes that satisfy the above design criterion for full-diversity using Generalized Butson-Hadamard matrices defined as follows:
\begin{definition}
A Generalized Butson-Hadamard (GBH) matrix \cite{Hor} is a $T \times T$ matrix $\textbf{M}$ with entries such that 
\begin{equation*}
\textbf{M}\textbf{M}^{H} = \textbf{M}^{H}\textbf{M} = T\textbf{I}_{T}
\end{equation*}
and the conjugate of every entry  $m_{ij}$ of $\textbf{M}$ is its inverse $m_{ij}^{-1}$ i.e $m_{ij}^{*} = m_{ij}^{-1}.$ 
\end{definition} 
%%%%%%%%%%%%%%%%%%%Code construction%%%%%%%%%%%%%%%
\section{A non-differential scheme and codes using cyclic unitary matrix groups}
\label{sec3}
We introduce a non-differential encoding scheme using a cyclic unitary matrix group \cite{HoS} of diagonal matrices at the source and relay matrices constructed from GBH matrices \cite{OgH1}, \cite{Hor}. Appropriate ingredients required at the source and the relays are as follows:
%%%%%%%%%%%%%%%At the source%%%%%%%%%%%%%%%%%%%%%%%%%%%%%%%%%%%%
\subsection{At the Source}
The source node is equipped with a finite abelian group $\mathcal{U}$ of size $L$ consisting of $2R \times 2R$ diagonal unitary matrices and a $2R \times 1$ complex vector, $\textbf{x}$. Let the group $\mathcal{U}$ be written as a direct product of $K$ cyclic groups $\mathcal{U}_{\nu}$, each of order $L_{\nu}$ for $\nu = 1, 2, \cdots, K$ as,
\begin{equation}
\label{group_U}
\mathcal{U} = \mathcal{U}_{1} \times \mathcal{U}_{2} \times \cdots \mathcal{U}_{K}
\end{equation}
where $L = \prod_{\nu = 1}^{K} L_{\nu}.$ When the set of $L_{\nu}$'s for $\nu = 1, 2, \cdots, K$ are pairwise relatively prime, the group $\mathcal{U}$ will be cyclic. Using the mixed-radix notation $l = \left( l_{1}, l_{2}, \cdots, l_{K}\right)$, where each $1 \leq l_{\nu} \leq L_{\nu}$, we denote every element of $\mathcal{U}$ as
\begin{equation*}
\textbf{D}^{l} = \prod_{\nu = 1}^{K} \textbf{D}_{\nu}^{l_{\nu}}
\end{equation*}
where $\textbf{D}_{\nu}$ is a generator of the $\nu^{th}$ cyclic group $\mathcal{U}_{\nu}$  given by
{\footnotesize
\begin{equation}
\label{generators}
\textbf{D}_{\nu} =  \mbox{diag}\left\lbrace exp\left(\frac{j2\pi u_{1 \nu}}{L_{\nu}}\right), ~ exp\left(\frac{j2\pi u_{2 \nu}}{L_{\nu}}\right) \cdots ~ exp\left(\frac{j2\pi u_{2R \nu}}{L_{\nu}}\right) \right\rbrace.
\end{equation}
}

\noindent For each $\nu$, $u_{i \nu} \in \left\lbrace 1, 2,\cdots L_{\nu}-1 \right\rbrace$ for $i = 1, 2, \cdots, 2R$ have to be chosen appropriately to achieve full-diversity and to maximise coding gain.
\noindent The $i$-th diagonal element of $\textbf{D}^{l}$ is given by
\begin{equation*}
\exp\left( j 2 \pi \sum_{\nu = 1}^{K} \frac{u_{i \nu}l_{\nu}}{L_{\nu}}\right).
\end{equation*}
Hence, the source is equipped with a code $\{\mathcal{U}, \textbf{x} \}$ where $\textbf{x} = \frac{1}{\sqrt{2R}} [1~1~1~\cdots~1]^T.$
%%%%%%%%%%%%%%%At the relays%%%%%%%%%%%%%%%%%%%%%%%%%%%%%%%
\subsection{At the relays}
\begin{figure*}
\begin{equation}
\label{zth element}
\textbf{V}^{z}_{12} = R\left[ exp\left(j 2\pi \sum_{\nu = 1}^{K} \frac{l_{\nu}(u_{z\nu}-u_{(R + z)\nu})}{L_{\nu}}\right) + exp\left(j 2\pi \sum_{\nu = 1}^{K} \frac{\hat{l}_{\nu}(u_{z\nu}-u_{(R + z)\nu})}{L_{\nu}}\right) \right]
\end{equation}
\hrule
\end{figure*}
Let $\textbf{M}$ be the $R \times R$ GBH matrix, using which we construct a $2R \times R$ matrix $\mathbf{\Gamma}$ as $\mathbf{\Gamma} = \left[ \textbf{M}^{T} ~ \textbf{M}^{T} \right]^{T}$. The columns of $\mathbf{\Gamma}$ are used to construct diagonal unitary matrices for the relays as $\textbf{A}_{j} = \mbox{diag} \left( \mathbf{\Gamma}_{1j}, \mathbf{\Gamma}_{2j}, \cdots \mathbf{\Gamma}_{2Rj} \right)$ where $\mathbf{\Gamma}_{ij}$ is the $(ij)-$th element of $\mathbf{\Gamma}$ and $j=1,2, \cdots, R.$
%%%%%%%%%%%%%%%%%%%%%%%%%%%%%%%%%%%%%%%%%%%%%%%%%
\subsection{Encoding Scheme}
The source node maps $\mbox{log}_{2}L$ bits of information on to one of the $L$ matrices from $\mathcal{U}$ say, $\textbf{D}^{k}$ for some $k$ where $1 \leqslant k \leqslant  L$ and transmits the $2R \times 1$ vector $\textbf{D}^{k}\mathbf{x}$ to all the relays. Each relay performs linear processing on its received vector using the unitary matrix $\textbf{A}_{j}$ and transmits a $2R$ length vectors to the destination. The distributed space-time codeword is of the form,
\begin{eqnarray}
\label{ncdstc_codeword}
\textbf{S}_{k} &=& \left[ \textbf{A}_{1}\textbf{D}^{k}\textbf{x} ~~ \textbf{A}_{2}\textbf{D}^{k}\textbf{x} ~\cdots ~\textbf{A}_{R}\textbf{D}^{k}\textbf{x} \right] \\
 &=& \textbf{D}^{k}\left[ \textbf{A}_{1}\textbf{x} ~~ \textbf{A}_{2}\textbf{x} ~\cdots ~\textbf{A}_{R}\textbf{x} \right] \nonumber \\
 &=&  \frac{1}{\sqrt{2R}}\textbf{D}^{k}\mathbf{\Gamma} \nonumber
\end{eqnarray}
\noindent
where we have used the fact that $\textbf{D}^k$ and $\textbf{A}_j$ commute for all $j.$ It can be observed that $\textbf{S}_{k}^{H}\textbf{S}_{k} = \textbf{I}_{R}$.
\begin{definition}
\label{cdstbc} 
The collection $\mathcal{C}$ of $2R \times R$ codeword matrices shown below when $k$ runs from 1 to L
\begin{equation}
\label{code}
\mathcal{C} = \left\lbrace \frac{1}{\sqrt{2R}}\textbf{D}^{k}\mathbf{\Gamma} \right\rbrace
\end{equation}
is called a Non-differential Abelian DSTBC (NADSTBC) and if the group is cyclic, then it is called a Non-differential Cyclic Distributed Space-Time Block Code (NCDSTBC).
\end{definition}

%%%%%%%%%%%%%%%%%%%%%%%%%%%%%%%%%%%%%%%%%%%%%%%%%%
%%%%%%%%%%%%Design criteria%%%%%%%%%%%%%%%%%%%%%%%%%%%%%%%%%
\subsection{Design criteria for $\textbf{D}_{\nu}$}
In this subsection, we derive a criteria for choosing the generators of each of $\mathcal{U}_{\nu}$ such that a NADSTBC is fully diverse.
%%%%%%%%%%%%%%%%%%General theorem%%%%%%%%%%%%%%%%%%%%%%%
\begin{theorem}
\label{thm2}
The diversity order of a wireless relay network with $R$ relays using a NADSTBC is $R$ if and only if the generators $\textbf{D}_{\nu}$ in \eqref{generators} are chosen such that, for each $i = 1, 2, \cdots R,$
\begin{equation}
\label{condition_abelian}
\sum_{\nu = 1}^{K} \frac{(l_{\nu} - \hat{l}_{\nu}) (u_{i\nu}-u_{(R + i)\nu})}{L_{\nu}} \notin \mathbb{Z}
\end{equation}
for all pair $(l, \hat{l})$ such that $l \neq \hat{l}.$
\end{theorem}
%%%%%%%%%%%%%%%%%%%%%%%%%%%%%%%%
\begin{proof} Let $\frac{1}{\sqrt{2R}}\textbf{D}^{l}\mathbf{\Gamma}$, $\frac{1}{\sqrt{2R}}\textbf{D}^{\hat{l}}\mathbf{\Gamma} \in \mathcal{C}$. The $2R \times 2R$ matrix as in \eqref{concate} is given by,
\begin{equation*}
\textbf{S}_{l \hat{l}}= \frac{1}{\sqrt{2R}}\left[ \textbf{D}^{l}\mathbf{\Gamma} ~ \textbf{D}^{\hat{l}}\mathbf{\Gamma}\right].
\end{equation*}
                                                                                       
\noindent Since $|\textbf{S}^{H}_{l \hat{l}}\textbf{S}_{l \hat{l}}| = |\textbf{S}_{l \hat{l}}\textbf{S}^{H}_{l \hat{l}}|$, we consider the choice of $\textbf{D}_{\nu}$ for $\nu = 1, 2, \cdots K$ such that $|\textbf{S}_{l \hat{l}}\textbf{S}^{H}_{l \hat{l}}| \neq 0$ where 
                                                                                       
\begin{equation}
\label{justopse}
\textbf{S}_{l \hat{l}}\textbf{S}^{H}_{l \hat{l}} = \frac{1}{2R}\left[\begin{array}{cc}
\textbf{V}_{11}  & \textbf{V}_{12}\\
\textbf{V}_{21} & \textbf{V}_{22}\\
\end{array}\right]\\
\end{equation}
with $\textbf{V}_{11} , \textbf{V}_{12} ,\textbf{V}_{21}$ and $\textbf{V}_{22}$ are $R \times R$ diagonal matrices given by
\begin{equation*}
\textbf{V}_{11} = 2R~\textbf{I}_{R \times R};~~
\textbf{V}_{21} = \textbf{V}_{12}^{H};~~
\textbf{V}_{22} = 2R~\textbf{I}_{R \times R}
\end{equation*}
and for each $z$ from $1, 2, \cdots, R$ the $z$-th diagonal entry of $\textbf{V}_{12}$ is given in \eqref{zth element}.
\noindent 
\noindent Applying the result,
{\small
\begin{equation*}
\left |\left[\begin{array}{rr}
\textbf{A}  & \textbf{B}\\
\textbf{C} & \textbf{D}\\
\end{array}\right]\right | = |\textbf{A}|| \textbf{D} - \textbf{C}\textbf{A}^{-1}\textbf{B}|
\end{equation*}
}

\noindent where $\textbf{A} , \textbf{B} , \textbf{C}$ and $\textbf{D}$ are square matrices of same order and  $\textbf{A}^{-1}$ exists, on the matrix in \eqref{justopse} and using \eqref{zth element}, we get
\begin{equation*}
|\textbf{S}_{\hat{l}l}\textbf{S}_{\hat{l}l}^{H}|  = \left(\frac{1}{2R}\right)^{2R}|\textbf{V}_{11}||\textbf{V}_{22} - \textbf{V}_{21}\textbf{V}_{11}^{-1}\textbf{V}_{12}|
\end{equation*}
\begin{equation*}
 = \left(\frac{1}{2R}\right)^{R} |\mbox{diag}\left( \delta_{1}~ \delta_{2} ~\cdots ~\delta_{R} \right)|
\end{equation*}
where,
{\small
\begin{eqnarray*}
\delta_{i} = R\left[1 - \cos\left(2\pi \sum_{\nu = 1}^{K} \frac{(l_{\nu} - \hat{l}_{\nu}) (u_{i\nu}-u_{(R + i)\nu})}{L_{\nu}}\right)\right] 
\end{eqnarray*}
}

\noindent For $|\textbf{S}_{l \hat{l}}\textbf{S}^{H}_{l \hat{l}}|$ not to be zero, the product $\delta_{1} \delta_{2} \cdots \delta_{R}$ must be a nonzero value. For each $\delta_{i}$ not to be zero, $u_{i\nu}$ and $u_{(i + R)\nu}$ for all $\nu = 1, 2, \cdots, K$ has to be chosen such that
\begin{equation*}
\sum_{\nu = 1}^{K} \frac{(l_{\nu} - \hat{l}_{\nu}) (u_{i\nu}-u_{(R + i)\nu})}{L_{\nu}} \notin \mathbb{Z}
\end{equation*}
for all $l,\hat{l}$ such that $l \neq \hat{l}.$ It is clear that the above condition on $\textbf{D}_{\nu}$ is both sufficient and necessary.
\end{proof}
%%%%%%%%%%%%%%%%%%%%%%%%%%%%%%%%
\begin{corollary}
\label{coral2}
DSTBCs from non-cyclic abelian groups are not fully diverse for any choice of $\textbf{D}_{\nu}$.
\end{corollary}
\begin{proof}
Since each of the cyclic groups $\mathcal{U}_{\nu} \subseteq \mathcal{U}$, from the special case of Theorem \ref{thm2} with $K=1,$ every $\textbf{D}_{\nu}$ must satisfy the criteria,
\begin{equation*}
\gcd\left( u_{\nu i} - u_{\nu (i + R)}, L_{\nu}\right)  = 1  \mbox{ for } i = 1, \cdots R.
\end{equation*}
Since at least two of the $L_{\nu}$ share a factor and each $\textbf{D}_{\nu}$ satisfy the above criteria, condition in \eqref{condition_abelian} of Theorem \ref{thm2} will not be satisfied for at least one pair of values  $l$ and $\hat{l}$. Hence, non-cyclic abelian DSTBCs cannot provide full diversity when used in the proposed scheme. 
\end{proof}
Henceforth, throughout the paper, we will consider only NCDSTBCs. We explicitly state Theorem \ref{thm2} for NCDSTBC as a corollary as:
\begin{corollary}
\label{coral1}
Let $\mathcal{U}$ in \eqref{group_U} be a cyclic group with a generator $\textbf{D}$ given by
\begin{equation*}
\textbf{D} = \mbox{diag}\left\lbrace exp\left(j 2 \pi \frac{u_{1}}{L}\right), exp\left(j 2 \pi \frac{u_{2}}{L}\right) \cdots exp\left(j 2 \pi \frac{u_{2R}}{L}\right) \right\rbrace.
\end{equation*}
A necessary and sufficient condition on the choice of $\textbf{D}$ such the that diversity order of the wireless relay network is $R$, is given by
\begin{equation*}
\gcd(u_{i} - u_{R + i}, L ) = 1 \mbox{ for each } i = 1, 2 , \cdots R.
\end{equation*}
\end{corollary}
\begin{proof}
The result follows as a special case of Theorem \ref{thm2} for $K$ = 1.
\end{proof}

Specializing the proof of Theorem \ref{thm2} for the cyclic group case, the coding gain for NCDSTBC is given by
\begin{equation}
\label{coding_gain_metric}
\phi =  \min_{1 \leqslant l , \hat{l} \leqslant  L, l \neq \hat{l} }\prod_{i = 1}^{R}\left[1 - \cos\left(\frac{2\pi(l-\hat{l})(u_{i} - u_{i + R})}{L}\right)\right].
\end{equation}
\noindent Thus, the vector $\textbf{v}$ has to be chosen such that $\phi$ is maximized, where
\begin{equation}
\label{v-vector}
\textbf{v} = \left( u_{1}-u_{R +1}, u_{2}-u_{R +2}  \cdots u_{R}-u_{2R}\right).
\end{equation}
%%%%%%%%%%%%%%%%%%%%%%%%%%%%%%%%%%%%%%%%%%%%%
From the result of Theorem \ref{thm2} and Corollary \ref{coral1}, the performance of NCDSTBC is characterized by \eqref{coding_gain_metric} which is independent of the unitary matrices $\textbf{A}_{j}$ at the relays and  dependent only on the generator of the cyclic group. 
\subsection{On the choice of $\mathbf{\Gamma}$}
\begin{figure}
\centering
\includegraphics[width= 2.9in]{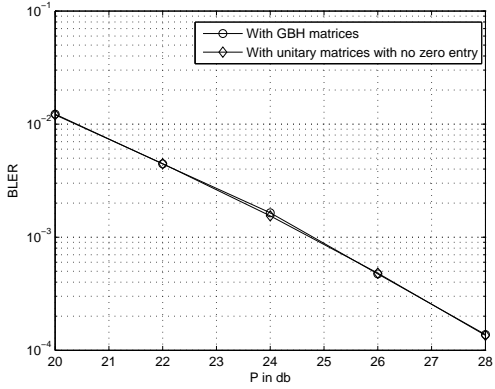}
\caption{Plot illustrating the independence of the BLER on relay specific diagonal matrices}
\label{GBH_normal-Unitary}
\end{figure}

Instead of a GBH matrix, $\textbf{M}$ can also be any $R \times R$ unitary matrix with no zero entry, using which $\mathbf{\Gamma}$ is constructed as
\begin{equation*}
\mathbf{\Gamma} = \sqrt{R}\left[\begin{array}{r}
\textbf{M}\\
\textbf{M}\\
\end{array}\right].
\end{equation*}
Though, the relay matrices $\textbf{A}_{j}$ are not unitary in this case, simulation results show that, for a given $\mathcal{U}$ at the source, BLER (Block error rate, which corresponds to errors in decoding a codeword is considered as error events of our interest throughout the paper) performance is independent of the above choice of $\textbf{M}$. This can be seen from Figure \ref{GBH_normal-Unitary} which is obtained using the code parameters $L =16$ and $\textbf{v} = [5, 7, 11, 1]$ for a network with 4 relays. However, decoding metric in \eqref{ML} doesn't hold in the latter case. Therefore, we continue to use only GBH matrices to construct $\textbf{A}_{j}$ in the rest of the paper. 
%%%%%%%%%%%%%%%Fig.3 begins %%%%%%%%%%%%%%%%%%%%%%%%%%%
%%%%%%%%%%%%%%%%%%%%%%%%%%%%%%%%%%%%%%%%%%%%%%
%%%%%%%%%%%%%%%%%%%ML decoding of our code%%%%%%%%%%%%%%%%
\section{Reduced ML Decoding complexity for NCDSTBCs and Node failures}
\label{sec4}
For a general class of unitary DSTBCs used in a partially-coherent scheme, an ML decoding metric is given in \eqref{ML}. In this section, we specialize this decoding metric for the class of  NCDSTBCs.
\begin{theorem} 
\label{thm3}
For a  NCDSTBC in \eqref{code}, the partially-coherent ML decoding given in \eqref{ML} reduces to 
\begin{equation}
\label{ML_cdstc}
arg\, \max_{k ~ \in ~ 1, 2, \cdots ~ L} \left(\textbf{y}_{1}^{H} \textbf{G}_{1}\textbf{y}_{1} + \mbox{Re} \left\lbrace{\textbf{y}_{1}^{H}\textbf{G}_{2}\textbf{y}_{2}}\right\rbrace + \textbf{y}_{2}^{H} \textbf{G}_{3}\textbf{y}_{2} \right).\\
\end{equation}
where,
\begin{equation*}
\textbf{y}_{1} = \left[y_{1} ~ y_{2} ~\cdots ~y_{R}\right]^{T},~
\textbf{y}_{2} = \left[y_{R +1} ~ y_{R + 2} \cdots ~y_{2R}\right]^{T} \mbox{ and }
\end{equation*}
$$\textbf{G}_{1} =  \mathbf{\Upsilon}_{1} \odot \mathbf{\Omega}, ~ \textbf{G}_{2} =  \mathbf{\Upsilon}_{2} \odot \mathbf{\Omega} \mbox{  and  } \textbf{G}_{3} =  \mathbf{\Upsilon}_{3} \odot \mathbf{\Omega}$$
where $\odot$ denotes the Hadamard product and $\mathbf{\Upsilon}_{1}, \mathbf{\Upsilon}_{2}$, $\mathbf{\Upsilon}_{3} \in \mathbb{C}^{R \times R}$ given by $\left[ \mathbf{\Upsilon}_{1}\right]_{i, j} = w^{(u_{i} - u_{j})k}$, $\left[ \mathbf{\Upsilon}_{2}\right]_{i, j} = w^{(u_{i} - u_{ R + j})k}$, $\left[ \mathbf{\Upsilon}_{3}\right]_{i, j} = w^{(u_{R + i} - u_{R + j})k}$ and
%{\scriptsize
%\begin{equation*}
%\mathbf{\Upsilon}_{1} = \left[\begin{array}{cccc}
%1 & w^{(u_{1} - u_{2})k} & \cdots & w^{(u_{1} - u_{R})k}\\
%w^{(u_{2} - u_{1})k} & 1 & \cdots & w^{(u_{2} - u_{R})k}\\
%\vdots & \vdots & \ddots & \vdots\\
%w^{(u_{R} - u_{1})k} & w^{(u_{R} - u_{2})k} & \cdots & 1\\
%\end{array}\right],
%\end{equation*}
%\begin{equation*}
%\mathbf{\Upsilon}_{2} = \left[\begin{array}{cccc}
%w^{(u_{1} - u_{R+1})k} & w^{(u_{1} - u_{R+2})k} & \cdots & w^{(u_{1} - u_{2R})k}\\
%w^{(u_{2} - u_{R+1})k} & w^{(u_{2} - u_{R+2})k} & \cdots & w^{(u_{2} - u_{2R})k}\\
%\vdots & \vdots & \ddots & \vdots\\
%w^{(u_{R} - u_{R+1})k} & w^{(u_{R} - u_{R+2})k} & \cdots & w^{(u_{R} - u_{2R})k}\\
%\end{array}\right],
%\end{equation*}
%\begin{equation*}
%\mathbf{\Upsilon}_{3} = \left[\begin{array}{cccc}
%1 & w^{(u_{R + 1} - u_{R + 2})k} & \cdots & w^{(u_{R + 1} - u_{2R})k}\\
%w^{(u_{R + 2} - u_{R + 1})k} & 1 & \cdots & w^{(u_{R + 2} - u_{2R})k}\\
%\vdots & \vdots & \ddots & \vdots\\
%w^{(u_{2R} - u_{R + 1})k} & w^{(u_{2R} - u_{R + 2})k} & \cdots & 1\\
%\end{array}\right]
%\end{equation*}
%}
$\mathbf{\Omega}$ is the $R \times R$ Hermitian matrix, with the $(i,j)-$th element given by
\begin{equation}
\label{new_omega}
\mathbf{\Omega}_{i,j} = \sum_{\lambda = 1}^{R} \beta_{\lambda}\left(m_{i \lambda}m_{j \lambda}^{*} \right) ~ \mbox{for} ~ i, j = 1, \cdots , R
\end{equation}
where $m_{ij}$ are the entries of $\textbf{M}$, the chosen GBH matrix and $\beta_{j} = |g_{j}|^{2}\left(|g_{j}|^{2} + \gamma \rho^{-1} \right)^{-1}~ \mbox{for} ~j = 1, \cdots , R.$
\end{theorem}
\begin{proof} The result follows by substituting \eqref{ncdstc_codeword} in \eqref{ML}.
\end{proof}

%%%%%%%%%%%%%%Choice of u fro reduced decoding complexity%%%%%%%%
\indent In the rest of this section, we present various choices on the vector $\textbf{v}$ in \eqref{v-vector} such that the computations required in decoding using the metric given by \eqref{ML_cdstc} is reduced. 
For large value of $P$, performance of the scheme is determined by the choice of the vector $\textbf{v}$. There are several ways of choosing the vector $\textbf{v}$ from various choices of $\textbf{u} = [u_{1}, u_{2} \cdots u_{2R}]$. The optimal choice of $\textbf{v}$ can still be made with appropriate values of $u_{1}, u_{2} \cdots u_{R}$ by keeping
\begin{equation*} 
\label{redu_cond_1}
u_{R +1} = u_{R + 2}  = u_{R + 3}  = \cdots = u_{2R}.
\end{equation*}
From the above choice of $v$, $\mathbf{\Upsilon}_{3}$ will be independent of the exponent $k$. Therefore the ML decoding metric in \eqref{ML_cdstc} reduces to
\begin{equation}
\label{ML_ncdstc_red}
arg\, \max_{k ~ \in ~ 1, 2, \cdots ~ L} \left(\textbf{y}_{1}^{\dagger} \textbf{G}_{1}\textbf{y}_{1} + \mbox{Re} \left\lbrace{\textbf{y}_{1}^{\dagger}\textbf{G}_{2}\textbf{y}_{2}}\right\rbrace  \right).\\
\end{equation}
Hence, compared to \eqref{ML_cdstc}, the number of computations required in \eqref{ML_ncdstc_red} to decode a codeword is reduced.\\
%However, for a NCDSTC with the choice of $v$ as in \eqref{redu_cond_1}, BLER at small and moderate values of $P$ will be worser than the NCDSTC with same $v$ constructed using $u$ which doesn't satisfy \eqref{redu_cond_1}.
%Simulation result illustrating the above behaviour is shown in Fig. \ref{low_snr}. The plot shows that the NCDSTC with $v$ = $\left[ 5, 5, 5, 5\right]$  using $u = \left[ 13, 11, 9, 7, 8, 6, 4, 2\right]$ performs better at low and moderate values of $P$ than the one with the same $v$ constructed using $u' = \left[7, 7, 7, 7, 2, 2, 2, 2\right]$. However, at large values of $P$, both codes have almost same BLER since both have the same $v$ as according to \eqref{high_snr_cond}. The difference in the BLER behaviour at lower values of $P$ between the two codes is due to a larger value of the metric in \eqref{low_snr_cond} for the former one compared to the latter.\\
%%%%%%%%%%%%%%%%%%%%%%%%%%%%%%%%%%%%%%%%%%%%%%%%%%%%%%%
%%%%%%%%%%%%%%%%%%%%%%%%%%%%%%%%%%%%%%%%%%%%%%
\indent The vector $\textbf{v}=(v_1,v_2,\cdots,v_{R})$ can also be chosen to be of the form
\begin{equation*}
\label{red_decod_vector_v}
v_{1} = v_{2} = \cdots = v_{R},
\end{equation*}
by choosing $u_{1} = u_{2} = \cdots = u_{R}$ and $u_{R +1} = u_{R + 2} = \cdots = u_{2R}$ such that $\gcd\left( {u_{1} - u_{R + 1}, L}\right)  = 1$.
In such a case, $\mathbf{\Upsilon}_{1}$ and $\mathbf{\Upsilon}_{3}$ are independent of the exponent $k$ and hence the decoding metric is written as
\begin{equation}
\label{reduced_ML}
arg\, \max_{k ~ \in ~ 1, 2, \cdots ~ L} \mbox{Re}~\left\lbrace{w^{(u_{1} -u_{3})k}\textbf{y}_{1}^{H}\mathbf{\Omega}\textbf{y}_{2}}\right\rbrace.
\end{equation}
In general, in order to estimate the most likely transmitted codeword from the source, the destination needs to perform matrix multiplication of order $R \times R$ according to \eqref{ML_cdstc} for every $k$ from 1 to $L$, where as, in the case of the metric given in \eqref{reduced_ML}, the destination needs to perform complex number multiplication for every $k$ to estimate the most likely transmitted codeword. Hence, the number of computations for decoding a codeword at the destination reduces substantially.
%%%%%%%%%%%%%%%%%%%%%The decoding metric given in \eqref{ML} has been used to carry out the simulations. 
\begin{figure}
\centering
\includegraphics[width= 2.9in]{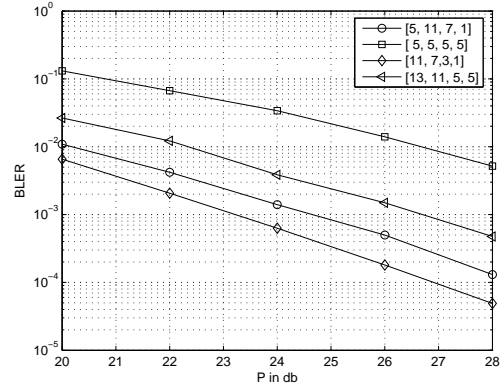}
\caption{BLER plot for 4 relays with L = 16 for different $\textbf{v}$}
\label{BLER_plot_for_several_vs}
\end{figure}

Figure \ref{BLER_plot_for_several_vs} shows the BLER performance against the total power, $P$ for different choices of generator $\textbf{D}$ with $L$ = 16 for a network with 4 relays. The plot shows that, for a choice of $\textbf{v}$ with $v_{1} = v_{2} = \cdots = v_{R}$, the NCDSTC loses out close to 5-6 db compared to the one with the choice of $\textbf{v}=[11, 7, 3, 1]$. The relay matrices used in all the simulations are $\textbf{A}_{1} = \textbf{I}_{8}$, $\textbf{A}_{2} = \mbox{diag}\left\lbrace 1, -1, 1, - 1, 1, -1, 1, -1\right\rbrace$, $\textbf{A}_{3} = \mbox{diag}\left\lbrace 1, 1, -1, - 1, 1, 1, -1, -1\right\rbrace$, $\mbox{ and }\textbf{A}_{4} = \mbox{diag}\left\lbrace 1, -1, -1, 1, 1, -1, -1, 1\right\rbrace$. The decoding metric given in \eqref{ML} has been used to carry out the simulations. In all our simulations, the total power $P$ has been distributed to the source node and the relay nodes as $P_{1} = \frac{P}{2}$ and $P_{2} = \frac{P}{2R}$.

%%%%%%%%%%%%%%%%%%%%%%%%%%%%%%%%%%%%%%%%%%%%%%
\subsection{Failure of relay nodes}

\begin{figure}
\centering
\includegraphics[width = 2.9in]{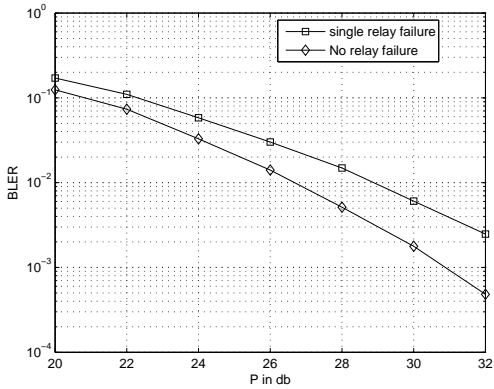}
\caption{BLER performance of NCDSTC with one relay down}
\label{relay_failure}
\end{figure}

In this subsection, we analyze the behavior of the proposed codes when a subset of the relay nodes fail to co-operate with the source i.e when a subset of the relays do not transmit information of the source to the destination. The structure of every codeword $\textbf{S}_{k}$ from the NCDSTC $\mathcal{C}$ is given in Definition \ref{cdstbc}. Similar to codes designed for non-differential non-coherent collocated MIMO systems, the columns of a codeword $\textbf{S}_{k}$ can be viewed as a basis of an $R$-dimensional subspace of $\mathbb{C}^{2R}$. Two codewords satisfy the full diversity condition means that, the $R$-dimensional subspace spanned by columns of each of them are non-intersecting. Therefore, a fully diverse NCDSTC can be viewed as a finite set of non-intersecting $R$- dimensional subspaces in  $\mathbb{C}^{2R}$. From the structure of the code, it is easy to see that, every relay contributes a basis element to each of the subspace (code-word). Let $\hat{\textbf{S}_{k}}$ denote a codeword, after say, $d$ of the $R$ relays fail to co-operate with the source. The columns of a new codeword $\hat{\textbf{S}_{k}}$ span a $(R - d)$ dimensional subspace of $R$-dimensional space spanned by the columns of $\textbf{S}_{k}$. Hence, the new set of subspaces (codewords) remains to be non-intersecting and the new code $\mathcal{\hat{C}}$ will remain to be fully diverse with diversity gain $R-d.$ Figure \ref{relay_failure} shows simulation results on the behavior of BLER when one of the relay fails in a network with 4 relays. The plot is obtained by using a code with parameters $L =16$ and $\textbf{v} = [11, 11, 11, 11]$.
%%%%%%%%%Fig. 5 begins%%%%%%%%%%%%%%%%

%%%%%%%%%%%%%%%%%%%%%%%%%%%%%%%%%%%%%%%%%
%%%%%%%%%%%%%%%%%%%%%%%%%%%%%%%%%%%%%%%%%%
\section{Discussion}
\label{sec5}
We considered the problem of code construction based on non-differential technique for wireless relay networks in a partially-coherent environment. Unitary distributed space time block codes were constructed using a cyclic group of diagonal matrices at the source and diagonal relay matrices constructed from GBH matrices. We have shown that the  BLER performance of the proposed scheme is independent of the relay specific unitary matrices. Further, a necessary and sufficient condition on the choice of the generator of cyclic group is provided for achieving full diversity and to minimize the PEP.  Various choices on the generator of cyclic group to reduce the ML decoding complexity at the destination was provided. The resistance of the proposed scheme to the failure of a subset of the relay nodes is also verified. It was also shown that, using non-cyclic abelian group of diagonal matrices at the source doesn't provide full diversity. An interesting direction for future work is to design non-differential encoding techniques such that the unitary distributed space time block codes can be constructed algebraically.

\end{document}